\documentclass[nofootinbib,twocolumn,preprintnumbers]{revtex4-1}
\usepackage{amsmath}
\usepackage{amssymb}
\usepackage{graphicx}
\usepackage{booktabs}
\usepackage{color}
\usepackage{hyperref}

\newcommand{\as}{\alpha_s}

\newcommand{\ptv}{p_{t}^{\tiny{\mbox{J,v}}}}

\newcommand{\ptvec}{{\vec{p}_t^{\,\tiny{\mbox{H}}}}}
\newcommand{\pt}{{p_t^{\tiny{\mbox{H}}}}}
\newcommand{\pthv}{{p_t^{\tiny{\mbox{H,v}}}}}
\newcommand{\ptj}{{p_t^{\tiny{\mbox{J}}}}}
\newcommand{\mh}{m_{\tiny{\mbox{H}}}}
\newcommand{\yh}{y^{\tiny{\mbox{H}}}}
\newcommand{\mur}{\mu_{\tiny{\mbox{R}}}}
\newcommand{\muf}{\mu_{\tiny{\mbox{F}}}}

\newcommand{\kto}{k_{t,1}}

\begin{document}

\title{Higgs transverse momentum with a jet veto: a double-differential resummation}

\preprint{CERN-TH-2019-101}

\author{Pier Francesco Monni$^1$, Luca Rottoli$^2$, Paolo Torrielli\,$^3$\vspace{1em}}

\affiliation{$^1$ CERN, Theoretical Physics Department, CH-1211 Geneva 23, Switzerland }
\affiliation{$^2$ Dipartimento di Fisica G. Occhialini, U2,
  Universit\`a degli Studi di Milano-Bicocca and INFN, Sezione di Milano-Bicocca, Piazza della Scienza, 3,
  20126 Milano, Italy}
\affiliation{$^3$ Dipartimento di Fisica and Arnold-Regge Center, Universit\`a di Torino and INFN, Sezione di Torino, Via P. Giuria 1, I-10125, Turin, Italy}

\begin{abstract}
  We consider the simultaneous measurement of the Higgs ($\pt$)
  and the leading jet ($\ptj$) transverse momentum in hadronic
  Higgs-boson production, and perform
  the resummation of the large logarithmic corrections that originate
  in the limit $\pt\,,\ptj \ll \mh$ up to
  next-to-next-to-leading-logarithmic order.
  This work constitutes the first simultaneous (double differential)
  resummation for two kinematic observables of which one involves a
  jet algorithm in hadronic collisions, and provides an important
  milestone in the theoretical understanding of joint resummations.
  As an application, we provide precise predictions for the Higgs
  transverse-momentum distribution with a veto $\ptj \leq \ptv$ on the
  accompanying jets, whose accurate description is relevant to the
  Higgs precision programme at the Large Hadron Collider.
\end{abstract}

\maketitle

The thorough scrutiny of the properties of the Higgs
boson~\cite{Aad:2012tfa,Chatrchyan:2012xdj} is central to the future
physics programme of the Large Hadron Collider (LHC).
In the High-Luminosity run of the LHC, the experimental
precision in Higgs-related measurements will increase significantly~\cite{Cepeda:2019klc},
hence allowing for detailed studies of the Higgs sector of
the Standard-Model (SM) Lagrangian.

A full exploitation of such measurements requires an unprecedented level
of precision in the theoretical description of the relevant observables.
In this context, a prominent role is played by kinematic distributions
of the Higgs boson and the accompanying QCD radiation, which are
sensitive to potential new-physics effects, such as modifications of
light-quark Yukawa couplings~\cite{Bishara:2016jga, Soreq:2016rae}, or
heavy new-physics
states~\cite{Banfi:2013yoa,Grojean:2013nya,Schlaffer:2014osa,Grazzini:2016paz,Banfi:2018pki,Banfi:2019xai}.
Experimental analyses of Higgs processes typically categorise the
collected events in {\it jet bins}, according to the different number
of jets --- collimated bunches of hadrons in the final state ---
produced in association with the Higgs boson.  Since the future
performance of the LHC will allow for the precise measurement of
kinematic distributions in different jet bins, it is paramount to
achieve an accurate theoretical understanding of Higgs observables at
the multi-differential level.

In this letter we consider Higgs-boson production in gluon fusion, the
dominant channel at the LHC, and we focus on the Higgs
transverse-momentum ($\pt$) spectrum in the presence of a veto $\ptv$
bounding the transverse momentum $\ptj$ of the hardest accompanying
jet.  Veto constraints of such a kind are customarily enforced
to enhance the Higgs signal with respect to its backgrounds,
relevant examples being the selection of $H \to W^+W^-$ events from
$t \bar t \to W^+W^- b \bar b$
production~\cite{Aaboud:2018jqu,Sirunyan:2018egh} or the
categorisation in terms of different initial
states~\cite{Aaboud:2018xdt}.

Fixed-order perturbative predictions of the $\pt$ spectrum in gluon
fusion are currently available at next-to-next-to-leading order (NNLO)
in the strong coupling
$\as$~\cite{Boughezal:2015dra,Boughezal:2015aha,Caola:2015wna,Chen:2016zka,Campbell:2019gmd}
in the infinite top-mass limit, and heavy-quark mass effects are known
up to next-to-leading order
(NLO)~\cite{Lindert:2017pky,Lindert:2018iug,Jones:2018hbb,Caola:2018zye,Neumann:2018bsx}.
 Fixed-order perturbation theory is, however, insufficient to accurately
 describe the observable considered here. When exclusive cuts on
 radiation are applied, it is well known that the convergence of the
 perturbative expansion is spoiled by the presence of logarithms
 $\ell \in \{\ln(\mh/\pt),\,\ln(\mh/\ptv)\}$ that become large in the
 limit $\pt, \ptv \ll \mh$, where the Higgs mass $\mh$ represents the
 typical hard scale of the considered process. In this regime, such large logarithmic
 terms must be summed to all perturbative orders to obtain a
 reliable theoretical prediction.
 The resummation accuracy is commonly defined at the level of the
 logarithm of the cumulative cross section, where terms of order
 $\as^n \ell^{n+1}$ are referred to as leading logarithms (LL),
 $\as^n \ell^n$ as next-to-leading logarithms (NLL),
 $\as^n \ell^{n-1}$ as next-to-next-to-leading logarithms (NNLL),
 and so on.
 The resummation of the inclusive $\pt$ spectrum has been carried out
 up to high perturbative
 accuracy~\cite{Bozzi:2005wk,Becher:2012yn,Neill:2015roa,Monni:2016ktx}
 and is currently known to N$^3$LL
 order~\cite{Bizon:2017rah,Chen:2018pzu}. Such calculations have been
 combined with NNLO fixed order in
 refs.~\cite{Bizon:2017rah,Chen:2018pzu,Bizon:2018foh} to obtain an
 accurate prediction across the whole $\pt$ spectrum.
 Similarly, the resummation of the jet-vetoed cross section has been
 achieved in
 refs.~\cite{Banfi:2012yh,Becher:2012qa,Banfi:2012jm,Becher:2013xia,Stewart:2013faa,Banfi:2013eda,Michel:2018hui},
 reaching NNLL accuracy matched to N$^3$LO~\cite{Banfi:2015pju}.
 Related resummations of the transverse momentum imbalance of the
 Higgs and the hardest jet have been also considered in
 refs.~\cite{Sun:2014lna,Sun:2016kkh,Chien:2019gyf}.
\sloppy

 In this work, we present the first {\it joint} resummation of
 both classes of logarithms, by obtaining a prediction which is
 differential in both $\pt$ and $\ptj$, and NNLL accurate in the limit
 $\pt, \ptj \ll \mh$. Specifically, we integrate the double-differential
 distribution $d\sigma/d\ptj d\pt$ over $\ptj$ up to
 $\ptj = \ptv$, which results in the single-differential $\pt$
 distribution with a jet veto.
 The results presented here are of phenomenological relevance in the
 context of the Higgs physics programme at the LHC, and constitute an
 important milestone in the theoretical understanding of the structure
 of resummations of pairs of kinematic observables, which has received
 increasing interest lately~\cite{Larkoski:2014tva,Procura:2018zpn,Lustermans:2019plv}.
 Different kinds of joint resummations for hadronic Higgs production
 have been considered in the literature. Relevant examples are
 combined resummations of logarithms of $\pt$ and
 small-$x$~\cite{Marzani:2015oyb,Forte:2015gve}, of $\pt$ and
 large-$x$~\cite{Laenen:2000ij,Kulesza:2003wn,Lustermans:2016nvk,Muselli:2017bad},
 of small-$x$ and large-$x$~\cite{Bonvini:2018ixe}, and of $\ptv$ and
 the jet radius~\cite{Banfi:2015pju}.

 To derive the main result of this letter, it is instructive to first
 consider the standard transverse-momentum
 resummation~\cite{Parisi:1979se,Collins:1984kg}, starting with a
 description of the effects that enter at NLL in a toy model with
 scale-independent parton densities.
 The core of the inclusive $\pt$ resummation lies in the description
 of soft, collinear radiation emitted off the initial-state gluons and
 strongly ordered in angle. Observing that in such kinematic
 configurations each emission is independent of the others, one
 obtains the following formula in impact-parameter ($b$) space
\begin{align}
  \label{eq:sigma-start}
\frac{d\sigma}{d^2\ptvec} & = \sigma_0 \int \frac{d^2\vec{b}}{4 \pi^2}
  e^{-i \vec{b} \cdot \ptvec}\notag\\
&\times\sum_{n=0}^\infty\frac{1}{n!}\prod_{i=1}^n\int [d k_i] M^2(k_i)
  \left(e^{i\vec{b}\cdot \vec{k}_{t,i}} - 1\right)\,,
\end{align}
where $\sigma_0$ denotes the Born cross section, and
$[d k_i] M^2(k_i)$ is the phase space and squared amplitude for
emitting a parton of momentum $k_i$.
The exponential factor in eq.~\eqref{eq:sigma-start} encodes in a
factorised form the kinematic constraint
$\delta^2(\ptvec - \sum_{i=1}^n \vec{k}_{t,i})$, while the $-1$ term
in the round brackets arises because, by unitarity, virtual
corrections come with a weight opposite to that of the real emissions,
but do not contribute to $\pt$. The factorisation of the phase-space
constraint allows for an exact exponentiation of the radiation in
eq.~\eqref{eq:sigma-start}, leading to the well known formula of
refs.~\cite{Parisi:1979se,Collins:1984kg}.

In order to include the constraint due to a veto on accompanying jets,
let us first consider the effect of a jet algorithm belonging to the
$k_t$-type family (such as the anti-$k_t$
algorithm~\cite{Cacciari:2008gp}). Owing to the strong angular
separation between the emissions, the clustering procedure at NLL will
assign each emission to a different jet~\cite{Banfi:2012yh}.
Therefore, imposing a veto $\ptv$ on the resulting jets corresponds to
constraining the real radiation with an extra factor
\begin{equation}
\label{eq:jet-veto}
\Theta(\ptv - \max\{k_{t,1},\dots,k_{t,n}\}) =
\prod_{i=1}^n\Theta(\ptv - k_{t,i})\,.
\end{equation}
Plugging the above equation into eq.~\eqref{eq:sigma-start} leads to
\begin{align}
  \label{eq:sigma-joint}
\frac{d\sigma(\ptv)}{d^2\ptvec} &= \sigma_0 \int \frac{d^2\vec{b}}{4 \pi^2}
  e^{-i \vec{b} \cdot \ptvec}\notag\\
&\hspace{-0.9cm}\times\sum_{n=0}^\infty\frac{1}{n!}\prod_{i=1}^n\int [d k_i] M^2(k_i)
  \left(e^{i\vec{b}\cdot \vec{k}_{t,i}} \Theta(\ptv - k_{t,i})-
  1\right)\notag\\
& = \sigma_0 \int \frac{d^2\vec{b}}{4 \pi^2}
  e^{-i \vec{b} \cdot \ptvec} e^{-S_{\rm NLL}}\,,
\end{align}
where the radiator $S_{\rm NLL}$ reads~\cite{Banfi:2012yh}
\begin{equation}
\label{eq:radiator}
S_{\rm NLL} = -\int [d k] M^2(k) \left(e^{i\vec{b}\cdot \vec{k}_{t}} \Theta(\ptv - k_{t})-
  1\right)\,.
\end{equation}
To evaluate the above integral, we can perform the integration over
the rapidity of the radiation $k$ and obtain
\begin{align}
  \label{eq:matrix-element}
\int [dk] M^2(k) =\int \frac{dk_{t}}{k_{t}} \frac{d\phi}{2\pi} R_{\rm NLL}^\prime(k_{t})\,,
\end{align}
with
\begin{align*}
 R_{\rm NLL}^\prime( k_{t}) = 4
    \left(\frac{\as^{\tiny{\mbox{CMW}}}(k_{t})}{\pi} C_A\ln \frac{\mh}{k_t}-\as(k_{t})\beta_0\right)\,,
\end{align*}
where $\beta_0$ is the first coefficient of the QCD beta function.
The coupling in the CMW scheme is defined
as~\cite{Catani:1990rr,Banfi:2018mcq,Catani:2019rvy}
$\as^{\tiny{\mbox{CMW}}}(k_{t})=\as(k_t) \left(1+
  \frac{\as(k_t)}{2\pi}\left[\left(\frac{67}{18}-\frac{\pi^2}{6}\right)C_A
    -\frac59 n_f\right]\right)$,
and includes the contribution of non-planar soft radiation necessary
for NLL accuracy in processes with two hard emitters.
The azimuthal integral of eq.~\eqref{eq:radiator} leads to
\begin{align}
\label{eq:rad_NLL_0}
S_{\rm NLL} &= -\int_0^{\mh} \frac{dk_{t}}{k_{t}} R_{\rm NLL}^\prime(k_{t})   \left(J_0(b k_t) -
  1\right) \notag\\
&+\int_0^{\mh} \frac{dk_{t}}{k_{t}} R_{\rm NLL}^\prime(k_{t})   J_0(b k_t) \Theta(k_{t} -\ptv)\,.
\end{align}
In the first integral, we exploit the large-$b$
property~\cite{Banfi:2012jm,Bizon:2017rah}
\begin{align}
\label{eq:J0-asympt}
 J_0(b k_t) \simeq 1 - \Theta(k_t - b_0/b) + {\cal O}({\rm N^3LL})\,,
\end{align}
with $b_0 = 2 e^{-\gamma_E}$, to recast eq.~\eqref{eq:rad_NLL_0} as
\begin{align}
\label{eq:rad_NLL}
S_{\rm NLL} =& - L g_1(\as L) - g_2 (\as L) \notag\\
& + \int_0^{\mh} \frac{dk_{t}}{k_{t}} R_{\rm NLL}^\prime(k_{t})   J_0(b k_t) \Theta(k_{t} -\ptv)\,,
\end{align}
where $\as \equiv \as(\mur)$ (with $\mur$ being the renormalisation
scale), $L=\ln(\mh b/b_0)$, and the $g_i$ functions are those used
in the standard $\pt$ resummation~\cite{SuppMaterial}.

The procedure that led to eq.~\eqref{eq:sigma-joint} can be used to
extend the above result to higher logarithmic orders.
The crucial observation is that, as already stressed, in impact-parameter
space the measurement function for $\pt$ is entirely factorised, resulting
in a phase factor $e^{i \vec{b}\cdot \vec{k}_t}$ for each emission $k$.
This implies that the jet-veto constraint $\Theta(\ptv - \ptj)$ can be
included by implementing the jet-veto resummation~\cite{Banfi:2012jm}
at the level of the $b$-space integrand, namely directly in impact-parameter
space.
We note incidentally that this observation can be applied to the
resummation of other pairs of observables for which the measurement
function can be factorised.

We now derive the NNLL result.
Starting from eq.~\eqref{eq:sigma-joint}, the first step is to promote
the $R_{\rm NLL}^\prime(k_{t}) $ function that appears in the radiator
$S_{\rm NLL}$ to NNLL. The corresponding expression is given in
refs.~\cite{Banfi:2012jm,Bizon:2017rah}, and leads to
\begin{align}
\label{eq:rad_NNLL}
S_{\rm NNLL} \equiv& \, - L g_1(\as L) - g_2(\as L) -
       \frac{\as}{\pi} g_3(\as L) \\
&+ \int_0^{\mh} \frac{dk_{t}}{k_{t}} R_{\rm NNLL}^\prime(k_{t})   J_0(b k_t) \Theta(k_{t} -\ptv)\notag\,.
\end{align}

The above step assumes that the veto on the radiation is encoded in a
phase-space constraint of the type~\eqref{eq:jet-veto}. While this
approximation is correct at NLL, where the jet algorithm does not
recombine the emissions with one another, it fails beyond this order.
Specifically, up to NNLL, at most two soft emissions can become close
in angle (three unordered soft emissions only contribute to N$^3$LL),
and therefore may get clustered into the same jet (whose momentum is
defined according to the so-called $E$-scheme, where the four momenta
of the constituents are added together). The configurations in which
the resulting cluster is the leading jet are not correctly described
by the constraint in~\eqref{eq:jet-veto}. In order to account for this
effect, one has to include a {\it clustering}
correction~\cite{Banfi:2012jm} in impact parameter space, that reads
\begin{eqnarray}
\label{eq:clust_def}
&&{\cal F}_{\rm clust} = \frac{1}{2!}\int [d k_a] [d k_b] M^2(k_a) M^2(k_b)
  J_{ab}(R) \,e^{i\vec{b}\cdot \vec{k}_{t,ab} }\nonumber\\
&&\hspace{3mm} \times \,\Big[\Theta(\ptv - k_{t,ab})-\Theta(\ptv - \max\{k_{t,a},k_{t,b}\})\Big]\,,
\end{eqnarray}
where $\vec{k}_{t,ab} = \vec{k}_{t,a} + \vec{k}_{t,b}$ and $k_{t,ab}$
is its magnitude. The constraint
$ J_{ab}(R)= \Theta\left(R^2-\Delta\eta_{ab}^2 - \Delta\phi_{ab}^2
\right)$
restricts the phase space to the region where the recombination
between the two emissions takes place. Here $R$ is the jet radius and
$\Delta \eta_{ab}$ and $\Delta\phi_{ab}$ are the pseudo-rapidity and
azimuthal separation between the two emissions, respectively.
We observe that eq.~\eqref{eq:clust_def} differs from the
corresponding clustering correction for the standard jet-veto
resummation~\cite{Banfi:2012jm} by the factor
$e^{i\vec{b}\cdot \vec{k}_{t,ab} }$, which accounts for the $\pt$
constraint in impact-parameter space.

Eq.~\eqref{eq:clust_def} describes the clustering correction due to
two independent soft emissions. A similar correction arises when the
two soft emissions $k_a$, $k_b$ are correlated, i.e. their squared
matrix element cannot be factorised into the product of two independent
squared amplitudes.
The contribution of a pair of correlated emissions is accounted for in
the CMW scheme for the strong coupling that was already used in the
NLL radiator~\eqref{eq:radiator}. However, such a scheme is obtained
by integrating {\it inclusively} over the correlated squared amplitude
$\tilde{M}^2(k_a,k_b)$, given in ref.~\cite{Dokshitzer:1997iz}.
While this inclusive treatment is accurate at NLL, at NNLL one needs
to correct for configurations in which the two correlated emissions
are not clustered together by the jet algorithm. This amounts to
including a {\it correlated} correction~\cite{Banfi:2012jm} of the form
\begin{eqnarray}
\label{eq:correl_def}
&&{\cal F}_{\rm correl} = \frac{1}{2!}\int [d k_a] [d k_b] \tilde{M}^2(k_a,k_b)
  (1-J_{ab}(R))e^{i\vec{b}\cdot \vec{k}_{t,ab} }\nonumber\\
&&\hspace{3mm} \times \,\Big[\Theta(\ptv - \max\{k_{t,a},k_{t,b}\}) - \Theta(\ptv - k_{t,ab})\Big]\,.
\end{eqnarray}
The corrections~\eqref{eq:clust_def} and~\eqref{eq:correl_def}
describe the aforementioned effects for a single pair of emissions. At
NNLL, all remaining emissions can be considered to be far in angle
from the pair $k_a$, $k_b$, and therefore they never get clustered
with the jets resulting from
eqs.~\eqref{eq:clust_def},~\eqref{eq:correl_def}.

As a final step towards a NNLL prediction, one must account for
non-soft collinear emissions off the initial-state particles. Since a
$k_t$-type jet algorithm never clusters the soft emissions discussed
above with non-soft collinear radiation, the latter can be
conveniently handled by taking a Mellin transform of the resummed
cross section. In Mellin space, the collinear radiation gives rise to
the scale evolution of the parton densities $f(\mu)$ and of the
collinear coefficient functions $C(\alpha_s)$. The latter, as well
as the hard-virtual corrections ${\cal H}(\alpha_s)$, must be included
at the one-loop level for a NNLL resummation. The equivalent of the
clustering and correlated corrections for hard-collinear radiation
enters only at N$^3$LL, and therefore is neglected in the following.

After applying to hard-collinear emissions the same procedure detailed
above for soft radiation, we obtain the main result of this letter,
namely the NNLL master formula for the $\pt$ spectrum with a jet veto
$\ptv$, differential in the Higgs rapidity $\yh$:
\begin{widetext}
\vspace{-5mm}
\begin{align}
\label{eq:sigma-joint-NNLL}
&\frac{d \sigma(\ptv)}{d \yh d^2\ptvec} = \frac{2\pi}{s}M^2_{ \rm g g\to {\tiny \mbox H}} \,{\cal
  H}(\alpha_s(\mh)) \,\int_{{\cal C}_1} \frac{d \nu_1}{2\pi i} \int_{{\cal C}_2} \frac{d \nu_2}{2\pi i} x_1^{-\nu_1}\,x_2^{-\nu_2} \int \frac{d^2\vec{b}}{4 \pi^2}
  e^{-i \vec{b} \cdot \ptvec}\,e^{-S_{\rm NNLL}} \left( 1 + {\cal F}_{\rm clust} + {\cal F}_{\rm
  correl}\right)  \\[5pt]
&~\times\!
  \left[{\cal P}\,e^{\int_{0}^{\mh}
  \frac{d \mu}{\mu}{\boldsymbol \Gamma}_{\nu_1}(\alpha_s(\mu) )\left(\Theta(\ptv-\mu)J_0(b \mu)-1\right)
  }\right]_{c_1 a_1}
  \!\left[{\cal P}\,e^{\int_{0}^{\mh}
  \frac{d \mu}{\mu}{\boldsymbol \Gamma}_{\nu_2}(\alpha_s(\mu)) \left(\Theta(\ptv-\mu)J_0(b \mu)-1\right)}\right]_{c_2 a_2}f_{\nu_1 ,a_1}(\mh)\,f_{\nu_2 ,a_2}(\mh)\notag \\[5pt]
&~ \times  e^{\int_{0}^{\mh}
  \frac{d \mu}{\mu}\left[{\boldsymbol
  \Gamma}^{(C)}_{\nu_1}(\alpha_s(\mu) )\right]_{g c_1}\left(\Theta(\ptv-\mu)J_0(b \mu)-1\right)
  } \,e^{\int_{0}^{\mh}
  \frac{d \mu}{\mu}\left[{\boldsymbol
  \Gamma}^{(C)}_{\nu_2}(\alpha_s(\mu))\right]_{g c_2}
  \left(\Theta(\ptv-\mu)J_0(b \mu)-1\right) }C_{\nu_1 ,g c_1}(\alpha_s(\mh)) \,C_{\nu_2 ,g c_2}(\alpha_s(\mh))\notag,
\end{align}
\vspace{-2mm}
\end{widetext}
where $x_{1,2} = \mh/\sqrt{s}\, e^{\pm \yh}$, and
$M^2_{ \rm g g\to {\tiny \mbox H}}$ is the Born squared matrix element
including the partonic flux factor. The $\nu_\ell$ subscripts denote
the Mellin transform, while the latin letters represent flavour
indices, and the sum over repeated indices is understood.
Here ${\boldsymbol \Gamma}_{\nu_\ell}$ and
${\boldsymbol \Gamma}^{(C)}_{\nu_\ell}$ are the anomalous dimensions
describing the scale evolution of the parton densities and coefficient
functions, respectively.
The contours ${\cal C}_1$ and ${\cal C}_2$ lie parallel to the
imaginary axis to the right of all singularities of the integrand. The
(anti-)path-ordering symbol ${\cal P}$ has a formal meaning, and
encodes the fact that the evolution operators are matrices in flavour
space. All the ingredients of eq.~\eqref{eq:sigma-joint-NNLL} are
given in ref.~\cite{SuppMaterial}.
The multi-differential distribution $d\sigma/d\ptj \, d\yh \, d^2\ptvec$ is
simply obtained by taking the derivative of
eq.~\eqref{eq:sigma-joint-NNLL} in $\ptv$.

All integrals entering the above formula are finite in four dimensions
and can be evaluated numerically to very high precision.
We point out that, similarly to the standard $\pt$
resummation~\cite{Monni:2016ktx,Bizon:2017rah}, the result in
eq.~\eqref{eq:sigma-joint-NNLL} can also be deduced directly in
momentum space, without resorting to an impact-parameter formulation.
The momentum-space approach is particularly convenient for
computational purposes, in that it gives access to differential
information on the QCD radiation, thereby enabling an efficient Monte
Carlo calculation. Therefore, we adopt the latter method for a
practical implementation of eq.~\eqref{eq:sigma-joint-NNLL}. The
relevant formulae are detailed in ref.~\cite{SuppMaterial}, and
implemented in the {\tt RadISH} program.

\begin{figure}[t!]
  \centering
  \includegraphics[width=0.9\linewidth]{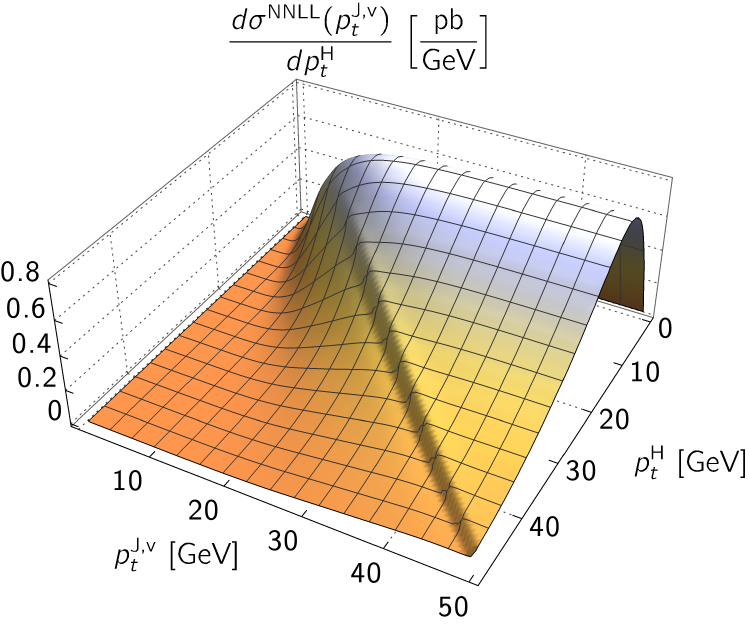}
  \caption{The NNLL differential
    distribution~\eqref{eq:sigma-joint-NNLL}, integrated over the
    Higgs-boson rapidity $\yh$ and over the $\ptvec$ azimuth, as a
    function of $\pt$ and $\ptv$. }
  \label{fig:cumulant_3d}
\end{figure}

For the numerical results presented below, we choose
$\sqrt{s} = 13\,{\rm TeV}$ and we adopt the {\tt NNPDF3.1}
set~\cite{Ball:2017nwa} of parton densities (PDFs) at NNLO, with
$\as(M_Z) = 0.118$. The evolution of the PDFs is performed with the
{\tt LHAPDF}~\cite{Buckley:2014ana} package and all convolutions are
handled with {\tt HOPPET}~\cite{Salam:2008qg}. We set the
renormalisation and factorisation scale to
$\mur=\muf=\mh = 125\,{\rm GeV}$, and
$R=0.4$. Figure~\ref{fig:cumulant_3d} shows
eq.~\eqref{eq:sigma-joint-NNLL} integrated over the rapidity of the
Higgs boson $\yh$ and over the $\ptvec$ azimuth, as a function of
$\pt$ and $\ptv$. We observe the typical peaked structure along the
$\pt$ direction, as well as the Sudakov suppression at small
$\ptv$. The two-dimensional distribution also features a Sudakov
shoulder along the diagonal $\pt \sim \ptv$, which originates from the
sensitivity of the differential spectrum to soft radiation in this
region beyond leading order~\cite{Catani:1997xc}.
Eq.~\eqref{eq:sigma-joint-NNLL} provides a resummation of the
logarithms associated with the shoulder in the regime
$\pt\sim \ptv \ll \mh$, which can be appreciated by the absence of an
integrable singularity in this region.

\begin{figure}[t!]
  \centering
  \includegraphics[width=0.9\linewidth]{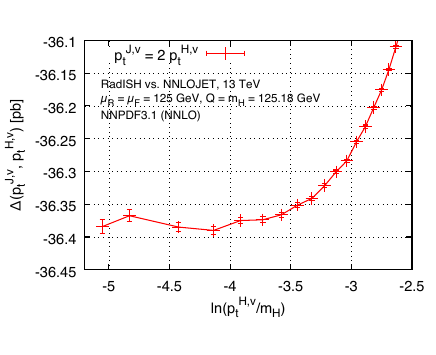}
  \caption{$\Delta(\ptv,\pthv)$, as defined in the text, at second
    order in $\as$ as a function of $\ln(\pthv/{m_{\rm H}})$, for
    $\ptv = 2\,\pthv$. This test features a slightly different Higgs
    mass, $\mh=125.18$ GeV.}
  \label{fig:Delta_check}
\end{figure}

To verify the correctness of eq.~\eqref{eq:sigma-joint-NNLL}, we
perform a number of checks. As a first observation, we note that in the
region $\ptv \gtrsim \mh$, the terms ${\cal F}_{\rm clust}$ and
${\cal F}_{\rm correl}$ vanish by construction and, as expected, one
recovers the NNLL resummation for the inclusive $\pt$ spectrum.
Conversely, considering the limit $\pt \gtrsim \mh$ (i.e.  small $b$),
eq.~\eqref{eq:sigma-joint-NNLL} reproduces the standard NNLL jet-veto
resummation of ref.~\cite{Banfi:2012jm} as detailed in
ref.~\cite{SuppMaterial}.
As a further test, we expand eq.~\eqref{eq:sigma-joint-NNLL} to second
order in $\as$ relative to the Born, and compare the result with an
${\cal O}(\alpha_s^2)$ fixed-order calculation for the inclusive
production of a Higgs boson plus one
jet~\cite{deFlorian:1999zd,Ravindran:2002dc,Glosser:2002gm}, with jets
defined according to the anti-$k_t$ algorithm~\cite{Cacciari:2008gp}.
In particular, to avoid the perturbative instability associated with
the Sudakov shoulder, we calculate the double cumulant
\begin{equation}
\sigma(\ptv,\pthv) \equiv \int d \yh d^2\ptvec
\frac{d \sigma(\ptv)}{d \yh d^2\ptvec} \Theta(\pthv - |\ptvec|)\notag\,,
\end{equation}
and define the quantity
\begin{equation}
\label{eq:Delta-def}
\Delta(\ptv,\pthv) = \sigma^{\rm NNLO}(\ptv,\pthv) -
\sigma^{\rm NNLL}_{\rm exp.}(\ptv,\pthv)\notag\,,
\end{equation}
where $\sigma^{\rm NNLO}(\ptv,\pthv)$ is computed by taking the
difference between the NNLO total Higgs-production cross
section~\cite{Harlander:2002wh,Anastasiou:2002yz,Ravindran:2003um},
obtained with the {\tt ggHiggs} program~\cite{Ball:2013bra}, and the
NLO Higgs+jet cross section for
$(\ptj > \ptv)\lor (\pt>\pthv )$, calculated with the {\tt
  NNLOJET} program~\cite{Chen:2016zka}. Given that the NNLL prediction
controls all divergent terms at the second perturbative order, one
expects the quantity $\Delta$ to approach a constant value of N$^3$LL
nature in the $\pt\to 0$ limit.
Figure~\ref{fig:Delta_check} displays this limit for
$\ptv = 2\,\pthv$, that shows an excellent convergence towards a
constant, thereby providing a robust test of
eq.~\eqref{eq:sigma-joint-NNLL}.

\begin{figure}[t!]
  \centering
  \includegraphics[trim={0 -0.2cm 0
    0},width=0.9\linewidth]{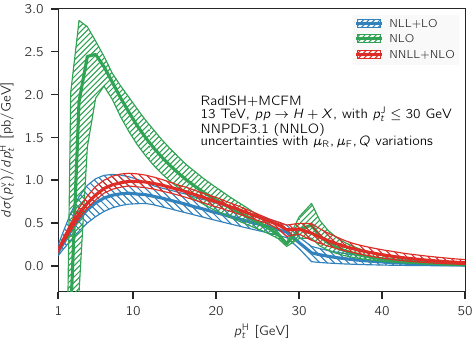}
  \caption{Matched NNLL+NLO (red band), NLL+LO (blue band), and
    fixed-order NLO (green band) $\pt$ differential distributions for
    $\ptv=30$ GeV, with theoretical uncertainties estimated as
    explained in the main text.}
  \label{fig:matched_JV30}
\end{figure}

As a phenomenological application of our result, we set
$\ptv = 30\,{\rm GeV}$ in accordance with the LHC experiments.
While eq.~\eqref{eq:sigma-joint-NNLL} provides an accurate description
of the spectrum in the small-$\pt$ region, in order to reliably extend
the prediction to larger $\pt$ values one needs to match the resummed
formula to a fixed-order calculation, in which the hard radiation is
correctly accounted for. We thus match the NNLL result to the NLO
Higgs+jet $\pt$ distribution obtained with the program {\tt
  MCFM-8.3}~\cite{Campbell:2015qma,Boughezal:2016wmq} by means of the
multiplicative matching formulated
in~\cite{Caola:2018zye,Bizon:2018foh,Bizon:2019zgf}.
We adopt the setup outlined above, and in addition we introduce the
resummation scale $Q$ as detailed in ref.~\cite{SuppMaterial} as a
mean to assess the uncertainties due to missing higher logarithmic
corrections. To estimate the theoretical uncertainty of our final
prediction, we perform a variation of the renormalisation and
factorisation scales by a factor of two about the central value
$\mur=\muf=\mh$, while keeping $1/2\leq \mur/\muf \leq 2$. Moreover,
for central $\mur$ and $\muf$ scales, we vary the resummation scale by
a factor of two around $Q=\mh/2$, and take the envelope of all
the above variations.
Figure~\ref{fig:matched_JV30} compares the NNLL+NLO prediction to the
NLL+LO, and to the fixed-order NLO result.
The integral of the NNLL+NLO (NLL+LO) distribution yields the
corresponding jet-vetoed cross section at NNLL+NNLO
(NLL+NLO)~\cite{Banfi:2012jm}.

We observe a good perturbative convergence for the resummed
predictions to the left of the peak, where logarithmic corrections
dominate. Above $\pt \sim 10\,{\rm GeV}$, the NNLL+NLO prediction
differs from the NLL+LO due to the large NLO $K$ factor in the
considered process. The residual perturbative uncertainty in the
NNLL+NLO distribution is of ${\cal O}(10\%)$ for $\pt \lesssim \ptv$.
The comparison to the NLO fixed order shows the importance of
resummation across the whole $\pt$ region, and a much reduced
sensitivity to the Sudakov shoulder\,\footnote{In
  Figure~\ref{fig:matched_JV30} we use a $2\,{\rm GeV}$ bin across the
  shoulder.} at $\pt\sim \ptv$.

In this letter we have formulated the first double-differential
resummation for an observable defined through a jet algorithm in
hadronic collisions. As a case study, we considered the production of
a Higgs boson in gluon fusion with transverse momentum $\pt$ in
association with jets satisfying the veto requirement
$\ptj \leq \ptv$.
In the limit $\pt, \ptv \ll \mh$, we performed the resummation of the
large logarithms $\ln(\mh/\pt),\,\ln(\mh/\ptv)$ up to NNLL, resulting
in an accurate theoretical prediction for this physical observable. As
a phenomenological application, we presented matched NNLL+NLO results
at the LHC.
Our formulation can be applied to the production of any colour-singlet
system, and it is relevant in a number of phenomenological
applications that will be explored in future work.\\

We would like to thank Andrea Banfi and Gavin Salam for stimulating
discussions on the subject of this letter, and Emanuele Re and Giulia
Zanderighi for constructive comments on the manuscript. We are very
grateful to Alexander Huss for kindly providing us with a cross check
of our results with the {\tt NNLOJET} program.
The work of PM has been supported by the Marie
Sk\l{}odowska Curie Individual Fellowship contract number 702610
Resummation4PS. LR is supported by the ERC Starting Grant REINVENT
(714788), and acknowledges the CERN Theoretical Physics Department
for hospitality and support during part of this work, and the CINECA
award under the ISCRA initiative for the availability of the high-performance
computing resources needed for this work.



\newpage

\onecolumngrid
\newpage
\appendix

\section*{Supplemental material}

\makeatletter
\renewcommand\@biblabel[1]{[#1S]}
\makeatother

\setcounter{figure}{0}

We here provide supplemental formulae that complete the discussions
and results of the letter.

\subsection{Explicit resummation formulae}

In the present section we report the explicit expressions for the
resummation functions $g_1$, $g_2$ and $g_3$ computed in
\cite{Bozzi:2005wk}. We report the results after the introduction of a
resummation scale $Q$, as described
in~\cite{Bozzi:2005wk,Banfi:2012jm}, that allows for an assessment of
the size of subleading logarithmic corrections.
With this convention, and a slight abuse of notation, we redefine
$L=\ln(Q b/b_0)$, and $\lambda=\as\beta_0L$.  Here $\as$ denotes
$\as(\mur)$, $Q$ is the resummation scale, of the order of the hard
scale $\mh$, while $\mur$ and $\muf$ denote the renormalisation and
factorisation scales, respectively. The Sudakov radiator $S$ then
reads (the formulae in the letter correspond to setting $Q=\mh$)
\begin{align}
  S_{\rm N(N)LL} \equiv& \, R_{\rm N(N)LL}(L)  + \int_0^{Q} \frac{dk_{t}}{k_{t}}  R_{\rm
                       N(N)LL}^\prime(k_{t})   J_0(b k_t) \Theta(k_{t} -\ptv)\,,
\end{align}
with
\begin{align}
R_{\rm NLL}(L) &= - L g_1(\as L) - g_2(\as L) ,\,\notag\\
R_{\rm NNLL}(L) &= R_{\rm NLL}(L) - \frac{\as}{\pi}
g_3(\as L)\,,
\end{align}
and
\begin{equation}
R_{\rm N(N)LL}'(k_t) \equiv \frac{d R_{\rm N(N)LL}(\ln(Q/k_t))}{d\,\ln(Q/k_t)}\,.
\end{equation}
The $g_i$ functions read
\begin{subequations}
\begin{align}
  g_{1}(\as L) &= \frac{A^{(1)}}{\pi\beta_{0}}\frac{2 \lambda +\ln (1-2 \lambda )}{2  \lambda }, \\
  g_{2}(\as L) &= \frac{1}{2\pi \beta_{0}}\ln (1-2 \lambda )
  \left(A^{(1)} \ln \frac{\mh^2}{Q^2}+B^{(1)}\right)
  -\frac{A^{(2)}}{4 \pi ^2 \beta_{0}^2}\frac{2 \lambda +(1-2
    \lambda ) \ln (1-2 \lambda )}{1-2
    \lambda} \notag\\
  &+A^{(1)} \bigg(-\frac{\beta_{1}}{4 \pi \beta_{0}^3}\frac{\ln
    (1-2 \lambda ) ((2 \lambda -1) \ln (1-2 \lambda )-2)-4
    \lambda}{1-2 \lambda}-\frac{1}{2 \pi \beta_{0}}\frac{(2 \lambda(1
    -\ln (1-2 \lambda ))+\ln (1-2 \lambda ))}{1-2\lambda} \ln
    \frac{\mur^2}{Q^2}\bigg)\,,\\
 g_{3}(\as L) &= \left(A^{(1)} \ln\frac{\mh^2}{Q^2}+B^{(1)}\right)
    \bigg(-\frac{\lambda }{1-2 \lambda} \ln
   \frac{\mu _{R}^2}{Q^2}+\frac{\beta_{1}}{2 \beta_{0}^2}\frac{2 \lambda
   +\ln (1-2 \lambda )}{1-2 \lambda}\bigg)
   -\frac{1}{2 \pi\beta_{0}}\frac{\lambda}{1-2\lambda}\left(A^{(2)}
       \ln\frac{\mh^2}{Q^2}+B^{(2)}\right)\notag \\
   &-\frac{A^{(3)}}{4 \pi ^2 \beta_{0}^2}\frac{\lambda ^2}{(1-2\lambda )^2}
   +A^{(2)} \bigg(\frac{\beta_{1}}{4 \pi  \beta_{0}^3 }\frac{2 \lambda  (3
   \lambda -1)+(4 \lambda -1) \ln (1-2 \lambda )}{(1-2 \lambda
   )^2}-\frac{1}{\pi \beta_{0}}\frac{\lambda ^2 }{(1-2 \lambda )^2}\ln\frac{\mur^2}{Q^2}\bigg) \notag\\
   & +A^{(1)} \bigg(\frac{\lambda  \left(\beta_{0} \beta_{2} (1-3 \lambda
   )+\beta_{1}^2 \lambda \right)}{\beta_{0}^4 (1-2 \lambda)^2}
   +\frac{(1-2 \lambda) \ln (1-2 \lambda ) \left(\beta_{0} \beta_{2}
   (1-2 \lambda )+2 \beta_{1}^2 \lambda \right)}{2\beta_{0}^4 (1-2 \lambda)^2}
   +\frac{\beta_{1}^2}{4 \beta_{0}^4}
   \frac{(1-4 \lambda ) \ln ^2(1-2 \lambda )}{(1-2 \lambda)^2}\notag\\
   &-\frac{\lambda ^2 }{(1-2 \lambda
   )^2} \ln ^2\frac{\mur^2}{Q^2}
   -\frac{\beta_{1}}{2 \beta_{0}^{2}}\frac{(2 \lambda  (1-2 \lambda)+(1-4 \lambda) \ln (1-2 \lambda ))
   }{(1-2\lambda )^2}\ln\frac{\mur^2}{Q^2}\bigg).
\end{align}
\end{subequations}
The coefficients of the QCD beta function up to three loops read
\begin{eqnarray}
  \beta_0 &=& \frac{11 C_A - 2 n_f}{12\pi}\,,\qquad
  \beta_1 = \frac{17 C_A^2 - 5 C_A n_f - 3 C_F n_f}{24\pi^2}\notag\,,\\
  \beta_2 &=& \frac{2857 C_A^3+ (54 C_F^2 -615C_F C_A -1415 C_A^2)n_f
       +(66 C_F +79 C_A) n_f^2}{3456\pi^3}\,,
\end{eqnarray}
and, for Higgs-boson production in gluon fusion, the coefficients
$A^{(i)}$ and $B^{(i)}$ entering the above formulae
are~\cite{deFlorian:2001zd,Becher:2010tm} (in units of $\as/(2\pi)$)
\begin{align}
  A^{(1)} =& \,2 C_A,
  \notag\\
  \vspace{1.5mm}
  A^{(2)} =&
  \left( \frac{67}{9}-\frac{\pi ^2}{3} \right) C_A^2
  -\frac{10}{9} C_A n_f,
  \notag\\
  \vspace{1.5mm}
  A^{(3)} =&
   \left( -22 \zeta_3 - \frac{67 \pi^2}{27}+\frac{11 \pi^4}{90}+\frac{15503}{324} \right) C_A^3
  + \left( \frac{10 \pi^2}{27}-\frac{2051}{162} \right) C_A^2 n_f\notag\\
  &+ \left( 4 \zeta_3-\frac{55}{12} \right) C_A C_F n_f
  + \frac{50}{81} C_A n_f^2,
  \notag\\
  \vspace{1.5mm}
  B^{(1)} =&
  -\frac{11}{3} C_A + \frac{2}{3}n_f,
  \notag\\
  \vspace{1.5mm}
  B^{(2)} =&
  \left( \frac{11 \zeta _2}{6}-6 \zeta _3-\frac{16}{3} \right) C_A^2
  + \left( \frac{4}{3}-\frac{\zeta _2}{3} \right) C_A n_f
  + n_f C_F\,.
\end{align}

We finally report the expressions for the collinear coefficient
function $C(\alpha_s(\mu))$ and the hard-virtual term
$\mathcal{H}(\mu)$ in eq.~\eqref{eq:sigma-joint-NNLL}:
\begin{align}
C_{ij}(z,\alpha_s(\mu)) &= \delta(1-z)\delta_{ij} +
  \frac{\alpha_s(\mu)}{2\pi}\,C_{ij}^{(1)}(z) +{\cal O}(\alpha_s^2)\,,\\
\mathcal{H}(\mu) &= 1 + \frac{\alpha_s(\mu)}{2\pi} \mathcal{H}^{(1)}+{\cal O}(\alpha_s^2)\,,
\end{align}
where
\begin{align}
 C_{ij}^{(1)}(z) &= -P_{ij}^{(0),\epsilon}(z)-\delta_{ij}\delta(1-z)C_A \frac{\pi^2}{12}+P_{ij}^{(0)}(z)
  \ln{\frac{Q^2}{\muf^{2}}}, \\
 \mathcal{H}^{(1)} &= H^{(1)}- \left( B^{(1)}+\frac{A^{(1)}}{2}\ln{\frac{\mh^2}{Q^2}}\right)\ln{\frac{\mh^2}{Q^2}}
 + {\rm d}_{B}~2\pi\beta_{0}\ln{\frac{\mur^2}{\mh^{2}}}\,.
\end{align}
Here ${\rm d}_{B}$ is the $\as$ power of the LO cross section
(${\rm d}_{B}= 2$ for Higgs production).  The coefficient $H^{(1)}$
encodes the pure hard virtual correction to the leading-order process
$gg\to H$, and in the $\overline{\rm MS}$ scheme it is given by
\begin{align}
 H^{(1)} &= C_A\left(5+\frac{7}{6}\pi^2\right)-3C_F\,.
\end{align}

$P_{ij}^{(0),\epsilon}(z)$ is the $\mathcal{O}(\epsilon)$ term of the LO splitting function $P_{ij}^{(0)}(z)$:
\begin{subequations}
\begin{align}
 P_{qq}^{(0),\epsilon}(z) &= -C_F(1-z)\,,\\
 P_{gq}^{(0),\epsilon}(z) &= -C_F z\,,\\
 P_{qg}^{(0),\epsilon}(z) &= -z(1-z)\,,\\
 P_{gg}^{(0),\epsilon}(z) &= 0.
\end{align}
\end{subequations}
The anomalous dimensions ${\boldsymbol \Gamma}_{\nu_\ell}$ and
${\boldsymbol \Gamma}^{(C)}_{\nu_\ell}$ in
eq.~\eqref{eq:sigma-joint-NNLL} are defined as
\begin{align}
[{\boldsymbol \Gamma}_{\nu_\ell}(\alpha_s(\mu))]_{ij} &= \frac{\alpha_s(\mu)}{\pi}\int_0^1 dz\,z^{\nu_\ell -1
  }\hat{P}_{ij}(z,\alpha_s(\mu))\,,\\
[{\boldsymbol \Gamma}_{\nu_\ell}^{(C)}(\alpha_s(\mu))]_{ij} &=
                                                              2\beta(\alpha_s(\mu))\frac{d\ln
                                                              C_{\nu_\ell,ij}(\alpha_s(\mu))}{d
                                                              \alpha_s(\mu)}\,,
\end{align}
where
\begin{equation}
C_{\nu_\ell,ij} (\alpha_s(\mu))\equiv \int_0^1 dz\,z^{\nu_\ell - 1} C_{ij}(z,\alpha_s(\mu))\,,
\end{equation}
and $\hat{P}_{ij}$ is the perturbative expansion of the regularised
splitting function (see e.g. ref.~\cite{Ellis:1991qj}). Finally, we
report the explicit formulae for the clustering~\eqref{eq:clust_def}
and correlated~\eqref{eq:correl_def} corrections used in the main
result of the letter. We find

\begin{align}
\label{eq:clustering}
{\cal F}_{\rm clust} &= \frac{1}{2!}\int_{0}^{\infty} \frac{dk_{t,a}}{k_{t,a}} \frac{dk_{t,b}}{k_{t,b}} \int_{-\infty}^{\infty}d\Delta \eta_{a b}\int_{-\pi}^{\pi}\frac{d\Delta\phi_{a
b}}{2\pi} \left(2 C_{A}
\frac{\as(k_{t,b})}{\pi}\right) \left(4 C_A\frac{\as(k_{t,a})}{\pi}
                       \ln \frac{\mh}{k_{t,a}}\right)\, J_{ab}(R)\notag\\
&\times
    \!\bigg[
   \Theta\Big(\ptv-|\vec{k}_{t,a} + \vec{k}_{t,b}|\Big) - \Theta\Big(\ptv-\max\{k_{t,a},k_{t,b}\}\Big)
   \bigg]\,e^{i\vec{b}\cdot \vec{k}_{t,a} }e^{i\vec{b}\cdot
  \vec{k}_{t,b} }\notag\\
&= \frac{1}{2!}\int_{0}^{\infty} \frac{dk_{t,a}}{k_{t,a}} \frac{dk_{t,b}}{k_{t,b}} \int_{-\infty}^{\infty}d\Delta \eta_{a b}\int_{-\pi}^{\pi}\frac{d\Delta\phi_{a
b}}{2\pi} \,8 \,C^2_{A}\,
\frac{\as^2}{\pi^2}\frac{\ln (Q/k_{t,a})}{(1-2\beta_0\as\ln(Q/k_{t,a}))^2} \, J_{ab}(R)\notag\\
&\times
    \!\bigg[
   \Theta\Big(\ptv-|\vec{k}_{t,a} + \vec{k}_{t,b}|\Big) - \Theta\Big(\ptv-\max\{k_{t,a},k_{t,b}\}\Big)
   \bigg]\,e^{i\vec{b}\cdot \vec{k}_{t,a} }e^{i\vec{b}\cdot
  \vec{k}_{t,b} } + {\cal O}({\rm N^3LL})\,,
\end{align}
where in the last step we have introduced the resummation scale $Q$
and neglected corrections beyond NNLL.
Similarly, within the same approximation, for the correlated
corrections we find
\begin{align}
\label{eq:correlated}
{\cal F}_{\rm correl}  &=  \frac{1}{2!}\int_{0}^{\infty} \frac{dk_{t,a}}{k_{t,a}} \frac{dk_{t,b}}{k_{t,b}} \int_{-\infty}^{\infty}d\Delta \eta_{a b}\int_{-\pi}^{\pi}\frac{d\Delta\phi_{a
b}}{2\pi}  \,8 \,C^2_{A}\,
\frac{\as^2}{\pi^2}\frac{\ln (Q/k_{t,a})}{(1-2\beta_0\as\ln(Q/k_{t,a}))^2} \, (1-J_{ab}(R))\notag\\
&\times {\cal C}\left(\Delta \eta_{ab},\Delta \phi_{ab},\frac{k_{t,a}}{k_{t,b}}\right)
     \!\bigg[
    \Theta\Big(\ptv-\max\{k_{t,a},k_{t,b}\}\Big)  - \Theta\Big(\ptv-|\vec{k}_{t,a} + \vec{k}_{t,b}|\Big)
    \bigg]\,e^{i\vec{b}\cdot \vec{k}_{t,a} }e^{i\vec{b}\cdot \vec{k}_{t,b} }\,.
\end{align}
The function ${\cal C}$ is defined as the ratio of the correlated part
of the double-soft squared amplitude to the product of the two
single-soft squared amplitudes, namely
\begin{equation}
 {\cal C}\left(\Delta \eta_{ab},\Delta
   \phi_{ab},\frac{k_{t,a}}{k_{t,b}}\right) \equiv
 \frac{\tilde{M}^2(k_a,k_b)}{M^2(k_a)\,M^2(k_b)}\,.
\end{equation}
Adopting the parametrisation of ref.~\cite{Dokshitzer:1997iz} for the
amplitudes, we have
\begin{align}
  M^2(k) = 2 \,C\, W(k)\,,
  \quad \tilde{M}^2(k_a,k_b) = C\, C_A (2 S + H_{\rm g})
  + C \,n_f H_{\rm q} \,,
\end{align}
where $W(k) \equiv 2/k_{t}^2$, and $C=C_A$ for Higgs production. The
functions $S$, $H_{\rm g}$, and $H_{\rm q}$ are given in
eqs.~(2.4)-(2.6) of ref.~\cite{Dokshitzer:1997iz}. We point out that
the symmetry factor $1/2!$ in eq.~\eqref{eq:correlated} accounts for
the contribution from two identical gluons. Conversely, the
contribution describing the emission of a $q\bar{q}$ pair in the
squared amplitude encodes an extra factor of $2$ that cancels against
the symmetry factor in this case.

We conclude this section by observing that all of the above integrals
have a Landau singularity that must be regulated with some
non-perturbative procedure. Given that the divergence occurs at very
small values of the transverse momentum (much below $1\,{\rm GeV}$),
it does not affect the region of phenomenological relevance considered
in our results. Therefore, in our study, we simply set the result to
zero at the singularity and below.

\subsection{Momentum-space formulation and implementation in {\tt
    RadISH}}

The momentum-space formulation of
refs.~\cite{Monni:2016ktx,Bizon:2017rah} allows a more
differential description of the radiation with respect to the
impact-parameter-space formulation used in the letter.
The access to differential
information comes at the cost of less compact equations, that however
can be efficiently evaluated through a Monte Carlo method.
The versatility of the Monte Carlo implementation can be exploited
observing that the resummation for the two considered observables
($\pt$ and $\ptj$) features the same momentum-space radiator
$R_{\rm N(N)LL}$. As a result, the joint resummation can be achieved
by modifying the phase-space constraint with respect to the inclusive
$\pt$ result of ref.~\cite{Monni:2016ktx}, and by adding the
clustering and correlated corrections discussed in the main text.

The resummation is more easily formulated at the level of the
double-cumulative distribution, namely
\begin{equation}
\frac{d\sigma(\ptv,\pthv)}{d\yh} \equiv \int_{0}^{\pthv} d\pt \int_{0}^{\ptv}d\ptj
\frac{d\sigma}{d\yh d\pt d\ptj}\,,
\end{equation}
and in the following we report both the NLL and the NNLL results in
turn.

\subsubsection{NLL formula}
At NLL, the measurement function for the pair of observables under consideration
for a state with $n$ emissions reads
\begin{equation}
\label{eq:veto-ordered}
\Theta(\ptv - \max\{k_{t,1},\dots,k_{t,n}\}) \Theta(\pthv - |\vec{k}_{t,1}+\dots+\vec{k}_{t,n}|)\,.
\end{equation}

Following ref.~\cite{Monni:2016ktx}, we single out the emission with
the largest transverse momentum $k_{t,1}$, and express the NLL cross
section as
\begin{align}
\label{eq:sigma-NLL-radish}
\frac{d\sigma^{\rm NLL}(\ptv,\pthv)}{d\yh}
&=\int_0^{\ptv}\frac{d k_{t,1}}{k_{t,1}}
                            \frac{d\phi_1}{2\pi}\,\int d{\cal Z} \,\frac{d}{d\,{L_{t,1}}}\left[-e^{-R_{\rm NLL}(L_{t,1})}\,{\cal L}_{\rm NLL}(\muf e^{-L_{t,1}}) \right] \Theta\Big(\pthv-|\vec{k}_{t,1}+\dots+\vec{k}_{t,n+1}|\Big)\,,
\end{align}
where $ L_{t,1} \equiv \ln(Q/k_{t,1}) $, and the factor ${\cal L}_{\rm NLL}$ reads
\begin{equation}
\label{eq:lumi_NLL}
{\cal L}_{\rm NLL}(\mu) \equiv \frac{2\pi}{s}M^2_{ \rm g g\to {\tiny \mbox H}}\, f_g(\mu,x_1) f_g(\mu,x_2)\,,
\end{equation}
where we introduced the explicit $x$ dependence of the parton
densities for later convenience.
We also introduced the measure $d{\cal Z}$ defined as
\begin{equation}
\int d{\cal Z} \equiv \epsilon^{\hat{R}'(k_{t,1})}\sum_{n=0}^{\infty} \frac{1}{n!}
 \prod_{i=2}^{n+1}\int_{\epsilon
   \kto}^{k_{t,1}}\frac{d k_{t,i}}{k_{t,i}} \frac{d\phi_i}{2\pi}
 \hat{R}'(k_{t,1})\,,
\end{equation}
with $\epsilon\ll 1$ an infrared, constant, resolution parameter that
allows for a numerical evaluation of eq.~\eqref{eq:sigma-NLL-radish}
in four space-time dimensions. We stress that the dependence on
$\epsilon$ entirely cancels in eq.~\eqref{eq:sigma-NLL-radish} for
sufficiently small values: in practice we set $\epsilon = e^{-20}$.
We also introduced the quantity~\cite{Monni:2016ktx}
\begin{equation}
\hat{R}^\prime(k_{t,1}) \equiv - \frac{d}{d L_{t,1}} \left( L_{t,1}
  g_1(\as L_{t,1})\right) = 4 C_A\frac{\as}{\pi} \, \frac{L_{t,1}}{(1-2\beta_0\as
L_{t,1})}\,.
\end{equation}

\subsubsection{NNLL formula}
Following the discussion at NLL, a first contribution to the NNLL
cross section is given by the NNLL formula for inclusive $\pt$,
supplemented by the jet-veto constraint. This
reads~\cite{Monni:2016ktx}
\begin{align}
  &\frac{d\sigma_{\rm incl}^{\rm NNLL}(\ptv,\pthv)}{d\yh}=\int_0^{\ptv}\frac{d
    k_{t,1}}{k_{t,1}} \frac{d\phi_1}{2\pi}\,\int d{\cal
    Z}\Bigg\{
    \,\frac{d}{d\,{L_{t,1}}}\left[-e^{-R_{\rm NNLL}(L_{t,1})}{\cal
    L}_{\rm NNLL}\left (\muf e^{-L_{t,1}}\right)
    \right] \Theta\Big(\pthv-|\vec{k}_{t,1}+\dots+\vec{k}_{t,n+1}|\Big) \notag\\
& + e^{-R_{\rm NLL}(L_{t,1})}\hat{R}'(k_{t,1})\!\int_{0}^{k_{t,1}}\frac{d k_{t,s_1}}{k_{t,s_1}} \frac{d\phi_{s_1}}{2\pi}
  \left[  \left(
    \delta\hat{R}'(k_{t,1})+\hat{R}''(k_{t,1})
    \ln\frac{k_{t,1}}{k_{t,s_1}}\right) {\cal
    L}_{\rm NLL}\left(\muf e^{-L_{t,1}}\right) - \frac{d}{d\,{L_{t,1}}}{\cal
    L}_{\rm NLL}\left(\muf e^{-L_{t,1}}\right)
    \right]\notag\\
&      \times\bigg[\Theta\Big(\pthv-|\vec{k}_{t,1}+\dots+\vec{k}_{t,n+1} + \vec{k}_{t,s_1}|\Big)
      -
      \Theta\Big(\pthv-|\vec{k}_{t,1}+\dots+\vec{k}_{t,n+1}|\Big)\bigg]\Bigg\} \,,
\end{align}
where ${\cal L}_{\rm NNLL}$ is given by
\begin{align}
\label{eq:lumi_NNLL}
{\cal L}_{\rm NNLL}(\mu) &\equiv \,\frac{2\pi}{s}M^2_{ \rm g g\to {\tiny \mbox H}}\,\bigg[ f_g\!\left(\mu,x_1\right)
  f_g\!\left(\mu,x_2\right)\left(1+\frac{\as}{2\pi}{\cal H}^{(1)}\right) +\notag\\
&  + \frac{\as}{2\pi}\frac{1}{1-2\as \beta_0 L_{t,1}}\sum_{k}\bigg(\int_{x_1}^1\frac{dz}{z}
  C_{gk}^{(1)}(z) f_k\!\left(\mu, \frac{x_1}{z}\right) f_g\!\left(\mu, x_2\right) + \{x_1\,\leftrightarrow\,x_2\}\bigg)\bigg]\,.
\end{align}
Finally, we introduced
\begin{align}
	\delta \hat R' (\kto) & \equiv - \frac{d \,g_2(\as L_{t,1})}{d L_{t,1}}\,, \\
	\hat R{''} (\kto)& \equiv \frac{d \,\hat{R}^\prime(k_{t,1})}{d L_{t,1}} \,.
\end{align}
When considering $\sigma_{\rm incl}^{\rm NNLL}(\ptv,\pthv)$
we used the phase-space constraint of eq.~\eqref{eq:veto-ordered}.
As discussed in the letter, this measurement
function assumes that the emissions are widely separated in rapidity
and therefore do not get clustered by the jet algorithm. However, at
NNLL at most two soft emissions are allowed to get arbitrarily close
in angle and to get clustered into the same jet. Accounting for this type
of configurations led to the formulation of the clustering
(${\cal F}_{\rm clust}$) and correlated (${\cal F}_{\rm correl}$)
corrections in the main text. In the following we will formulate these
two corrections directly in momentum space.

The clustering correction can be expressed as
\begin{align}
&\frac{d\sigma_{\rm clust}^{\rm NNLL}(\ptv,\pthv)}{d\yh} = \int_0^{\infty}\frac{d
    k_{t,1}}{k_{t,1}} \frac{d\phi_1}{2\pi}\,\int d{\cal
    Z}\,e^{-R_{\rm NLL}(L_{t,1})}\,{\cal L}_{\rm
  NLL}\left(\muf e^{-L_{t,1}}\right) \,8 \,C^2_{A}\,
\frac{\as^2}{\pi^2}\frac{L_{t,1}}{(1-2\beta_0\as\,L_{t,1})^2}\,\Theta\left(\ptv
  - \max_{i > 1}\{k_{t,i}\}\right)\notag\\
&\,\times\Bigg\{\int_{0}^{k_{t,1}} \hspace{-0.3em}\frac{d
  k_{t,s_1}}{k_{t,s_1}} \frac{d \phi_{s_1}}{2\pi}\!
\int_{-\infty}^{\infty}  \hspace{-1em}d\Delta\eta_{1s_1} \,
  J_{1 s_1}(R)\bigg[
   \Theta\Big(\ptv-|\vec{k}_{t,1} + \vec{k}_{t,s_1}|\Big) -
  \Theta\Big(\ptv-k_{t,1}\Big) \bigg]\Theta\Big(\pthv-|\vec{k}_{t,1}+\dots+\vec{k}_{t,n+1} + \vec{k}_{t,s_1}|\Big)\notag\\
&\, + \frac{1}{2!}\hat{R}'(k_{t,1})\,\int_{0}^{k_{t,1}}\frac{d k_{t,s_1}}{k_{t,s_1}} \frac{d
  k_{t,s_2}}{k_{t,s_2}} \frac{d \phi_{s_1}}{2\pi} \frac{d \phi_{s_2}}{2\pi}
  \!\int_{-\infty}^{\infty}\hspace{-1em} d\Delta\eta_{s_1 s_2} \,
  J_{s_1 s_2}(R)\bigg[
   \Theta\Big(\ptv-|\vec{k}_{t,s_1}+ \vec{k}_{t,s_2}|\Big) -
  \Theta\Big(\ptv-\max\{k_{t,s_1},k_{t,s_2}\}\Big) \bigg]\notag\\
&\,\times\Theta\Big(\pthv-|\vec{k}_{t,1}+\dots+\vec{k}_{t,n+1} +
  \vec{k}_{t,s_1}+ \vec{k}_{t,s_2}|\Big)\,\Theta\left(\ptv - \kto\right)\Bigg\}\,,
\end{align}
where we have explicitly separated the configuration in which one of
the two clustered emissions is the hardest ($k_1$), from the
configuration in which both clustered emissions have
$k_{t,s_1/s_2} \leq k_{t,1}$. Although the latter step is not
necessary, we find it convenient to keep the two contributions
separate for a Monte Carlo implementation. The same arguments can be
applied to the correlated correction, which can be expressed as

\begin{align}
&\frac{d\sigma_{\rm correl}^{\rm NNLL}(\ptv,\pthv)}{d\yh} = \int_0^{\infty}\frac{d
    k_{t,1}}{k_{t,1}} \frac{d\phi_1}{2\pi}\,\int d{\cal
    Z}\,e^{-R_{\rm NLL}(L_{t,1})}\,{\cal L}_{\rm
  NLL}\left(\muf e^{-L_{t,1}}\right) \,8 \,C^2_{A}\,
\frac{\as^2}{\pi^2}\frac{L_{t,1}}{(1-2\beta_0\as\,L_{t,1})^2}\,\Theta\left(\ptv
  - \max_{i > 1}\{k_{t,i}\}\right)\notag\\
&\,\times\Bigg\{\,\int_{0}^{k_{t,1}}\frac{d
  k_{t,s_1}}{k_{t,s_1}} \frac{d \phi_{s_1}}{2\pi}
\!  \int_{-\infty}^{\infty}\hspace{-1em}d\Delta\eta_{1 s_1} \,{\cal C}\left(\Delta \eta_{1 s_1},\Delta \phi_{1 s_1},\frac{k_{t,1}}{k_{t,s_1}}\right)\,
 \left(1- J_{1 s_1}(R)\right)\notag\\
&\,\times\bigg[
  \Theta\Big(\ptv-k_{t,1}\Big) - \Theta\Big(\ptv-|\vec{k}_{t,1} + \vec{k}_{t,s_1}|\Big)\bigg]\Theta\Big(\pthv-|\vec{k}_{t,1}+\dots+\vec{k}_{t,n+1} + \vec{k}_{t,s_1}|\Big)\notag\\
&\, + \frac{1}{2!}\hat{R}'(k_{t,1})\,\int_{0}^{k_{t,1}}\frac{d k_{t,s_1}}{k_{t,s_1}} \frac{d
  k_{t,s_2}}{k_{t,s_2}} \frac{d \phi_{s_1}}{2\pi} \frac{d \phi_{s_2}}{2\pi}
  \!\int_{-\infty}^{\infty}\hspace{-1em}d\Delta\eta_{s_1 s_2} \,{\cal C}\left(\Delta \eta_{s_1 s_2},\Delta
  \phi_{s_1 s_2},\frac{k_{t,s_2}}{k_{t,s_1}}\right) \left(1-
  J_{s_1 s_2}(R)\right) \,\Theta\left(\ptv - \kto\right)\notag\\
&\,\times\bigg[
   \Theta\Big(\ptv-\max\{k_{t,s_1},k_{t,s_2}\}\Big) -\Theta\Big(\ptv-|\vec{k}_{t,s_1}+ \vec{k}_{t,s_2}|\Big) \bigg]\Theta\Big(\pthv-|\vec{k}_{t,1}+\dots+\vec{k}_{t,n+1} + \vec{k}_{t,s_1}+ \vec{k}_{t,s_2}|\Big)\Bigg\}\,.
\end{align}

The NNLL double-cumulative distribution is then obtained by summing
the three contributions, namely
\begin{equation}
\sigma^{\rm NNLL}(\ptv,\pthv) = \sigma_{\rm incl}^{\rm NNLL}(\ptv,\pthv)
+ \sigma_{\rm clust}^{\rm NNLL}(\ptv,\pthv) + \sigma_{\rm correl}^{\rm NNLL}(\ptv,\pthv) \,.
\end{equation}
We refer to Section 4.3 of ref.~\cite{Bizon:2017rah} for the Monte
Carlo evaluation of the above equations, and to Section 4.2 of the
same article for the procedure used to expand them at a fixed
perturbative order.

\subsection{Asymptotic limits of the joint-resummation formula}
In this section we perform the asymptotic limits of
eq.~\eqref{eq:sigma-joint-NNLL}, and verify that it reproduces the
NNLL results for $\pt$ and jet-veto resummation, respectively. We
start by taking the limit $\ptv \sim \mh \gg \pt$. Using the fact that
\begin{align}
S_{\rm NNLL} \underset{{\ptv \sim \,\mh \gg \,\pt}}{\sim} R_{\rm NNLL}(L) \,,
\end{align}
and observing that eqs.~\eqref{eq:clustering},~\eqref{eq:correlated}
vanish since both $\Theta$ functions are satisfied, we obtain
\begin{align}
\frac{d \sigma(\ptv)}{d \yh d^2\ptvec} &\simeq\frac{2\pi}{s} M^2_{ \rm g g\to {\tiny \mbox H}} \,{\cal
  H}(\alpha_s(\mh)) \,\int_{{\cal C}_1} \frac{d \nu_1}{2\pi i} \int_{{\cal C}_2} \frac{d \nu_2}{2\pi i} x_1^{-\nu_1}\,x_2^{-\nu_2} \int \frac{d^2\vec{b}}{4 \pi^2}
  e^{-i \vec{b} \cdot \ptvec}\,e^{-R_{\rm NNLL}} \\
&~\times
  f_{\nu_1 ,a_1}(b_0/b)\,f_{\nu_2 ,a_2}(b_0/b)
C_{\nu_1 ,g a_1}(\alpha_s(b_0/b)) \,C_{\nu_2 ,g a_2}(\alpha_s(b_0/b)) \notag,
\end{align}
that, upon performing the Mellin integrals, coincides with the
inclusive $\pt$ resummation (see for instance
ref.~\cite{Bozzi:2005wk}).

Similarly, we now consider the limit $\pt \sim \mh \gg \ptv$. This
limit corresponds to taking the impact parameter $b$ to zero while
keeping $b \,\pt$ fixed. We observe that this limit probes the region in
which the approximation~\eqref{eq:J0-asympt} cannot be made. This
issue is commonly circumvented by modifying the $b$-space logarithms
as in~\cite{Bozzi:2005wk}. Alternatively, one can avoid making the
approximation~\eqref{eq:J0-asympt} in the first place, which
guarantees the $b\to 0$ limit to be well defined. In this case one
exploits the fact that
\begin{equation}
\lim_{b\to 0} J_0(b x) = 1\,,
\end{equation}
and obtains
\begin{align}
\label{eq:ptv-limit}
\frac{d \sigma(\ptv)}{d \yh d^2\ptvec} &\simeq \frac{2\pi}{s}M^2_{ \rm g g\to {\tiny \mbox H}} \,{\cal
  H}(\alpha_s(\mh)) \,\int_{{\cal C}_1} \frac{d \nu_1}{2\pi i} \int_{{\cal C}_2} \frac{d \nu_2}{2\pi i} x_1^{-\nu_1}\,x_2^{-\nu_2} \int \frac{d^2\vec{b}}{4 \pi^2}
  e^{-i \vec{b} \cdot \ptvec}\,e^{-S_{\rm NNLL}} \left( 1 + {\cal F}_{\rm clust} + {\cal F}_{\rm
  correl}\right)\notag  \\
&~\times\!
  \left[{\cal P}\,e^{-\int_{\ptv}^{\mh}
  \frac{d \mu}{\mu}{\boldsymbol \Gamma}_{\nu_1}(\alpha_s(\mu) )
  }\right]_{c_1 a_1}
  \!\left[{\cal P}\,e^{-\int_{\ptv}^{\mh}
  \frac{d \mu}{\mu}{\boldsymbol \Gamma}_{\nu_2}(\alpha_s(\mu)) }\right]_{c_2 a_2}f_{\nu_1 ,a_1}(\mh)\,f_{\nu_2 ,a_2}(\mh)\notag\\
&~ \times \! \,e^{-\int_{\ptv}^{\mh}
  \frac{d \mu}{\mu}\left[{\boldsymbol
  \Gamma}^{(C)}_{\nu_1}(\alpha_s(\mu) )\right]_{g c_1} 
  }\!\,e^{-\int_{\ptv}^{\mh}
  \frac{d \mu}{\mu}\left[{\boldsymbol
  \Gamma}^{(C)}_{\nu_2}(\alpha_s(\mu))\right]_{g c_2}
 }C_{\nu_1 ,g c_1}(\alpha_s(\mh)) \,C_{\nu_2 ,g c_2}(\alpha_s(\mh))\notag\\
& = \frac{2\pi}{s}M^2_{ \rm g g\to {\tiny \mbox H}} \,{\cal
  H}(\alpha_s(\mh)) \,\int_{{\cal C}_1} \frac{d \nu_1}{2\pi i} \int_{{\cal C}_2} \frac{d \nu_2}{2\pi i} x_1^{-\nu_1}\,x_2^{-\nu_2} \int \frac{d^2\vec{b}}{4 \pi^2}
  e^{-i \vec{b} \cdot \ptvec}\,e^{-S_{\rm NNLL}} \left( 1 + {\cal F}_{\rm clust} + {\cal F}_{\rm
  correl}\right)\notag  \\
&~\times\!
  f_{\nu_1 ,a_1}(\ptv)\,f_{\nu_2 ,a_2}(\ptv)
 C_{\nu_1 ,g a_1}(\alpha_s(\ptv)) \,C_{\nu_2 ,g a_2}(\alpha_s(\ptv)) \,,
\end{align}
where in the second line we applied the evolution operators from $\mh$
to $\ptv$ to the parton densities and coefficient functions. Finally, by
observing that
\begin{align}
S_{\rm NNLL} \underset{{\pt \sim \,\mh \gg \,\ptv}}{\sim} R_{\rm NNLL}(\ln(\mh/\ptv)) \,,
\end{align}
the integral over the impact parameter $b$ becomes trivial
\begin{equation}
\int \frac{d^2\vec{b}}{4 \pi^2}
  e^{-i \vec{b} \cdot \ptvec} = \delta^2(\ptvec)\,.
\end{equation}
As a consequence, upon integration over $\ptvec$,
eq.~\eqref{eq:ptv-limit} yields
\begin{align}
\frac{d \sigma(\ptv)}{d \yh} &= \frac{2\pi}{s}M^2_{ \rm g g\to {\tiny \mbox H}} \,{\cal
  H}(\alpha_s(\mh)) \,e^{-R_{\rm NNLL}(\ln(\mh/\ptv))} \left( 1 + {\cal F}_{\rm clust} + {\cal F}_{\rm
  correl}\right)\notag  \\
&~\times\!
  \left[f(\ptv)\otimes C(\alpha_s(\ptv))\right]_g (x_1)\left[f(\ptv)\otimes C(\alpha_s(\ptv))\right]_g (x_2)\,,
\end{align}
that coincides with the standard jet-veto
resummation~\cite{Banfi:2012jm} differential in the Higgs-boson
rapidity, where the convolution between two functions $f(x)$ and
$g(x)$ is defined as
\begin{equation}
[f\otimes g](x) \equiv \int_x^1 \frac{d z}{z} f(z)
g\left(\frac{x}{z}\right)\,.
\end{equation}

\end{document}